\documentclass[tighten,iop]{emulateapj}

\usepackage[english]{babel}
\usepackage{amsmath}
\usepackage{amsfonts}
\usepackage{amssymb}
\usepackage{graphicx}
\usepackage[dvips]{color} 

\usepackage{natbib}

\shorttitle{}
\shortauthors{Nagakura et al.}

\begin{document}

\title{Semi-dynamical approach to the shock revival in core-collapse supernovae}

\author{Hiroki Nagakura$^{1,2}$, Yu Yamamoto$^{3}$ and Shoichi Yamada$^{2,3}$}
\address{$^1$Yukawa Institute for Theoretical Physics, Kyoto
  University, Oiwake-cho, Kitashirakawa, Sakyo-ku, Kyoto, 606-8502,
  Japan}
\address{$^2$Advanced Research Institute for Science \&
Engineering, Waseda University, 3-4-1 Okubo,
Shinjuku, Tokyo 169-8555, Japan}
\address{$^3$Department of Science and Engineering, Waseda
  University, 3-4-1 Okubo, Shinjuku, Tokyo 169-8555, Japan}
%\email{hiroki@heap.phys.waseda.ac.jp}
%\address{}

\begin{abstract}
We develop a new semi-dynamical method to study shock revival by neutrino heating in core-collapse supernovae. 
Our new approach is an extension of the previous studies that employ spherically symmetric, steady, shocked accretion flows together with the light bulb approximation. The latter has been 
widely used in the supernova community for the phenomenological investigation of the criteria for successful supernova explosions. 
In the present approach, on the other hand, we get rid of the steady-state condition and take into account shock wave motions instead. 
We have in mind the scenario that not the critical luminosity but the critical fluctuation generated by hydrodynamical 
instabilities such as SASI and neutrino-driven convection in the post-shock region determines the onset 
of shock revival.
 After confirming that the new approach indeed captures 
the dynamics of revived shock wave qualitatively, we then apply the method to various initial conditions and find that there
is a critical fluctuation for shock revival, which can be well fit by the following formula: 
$f_{crit} \sim 0.8 \times (M_{in}/1.4M_{\odot}) \times \{ 1 - (r_{sh}/10^{8} {\rm cm})  \}$, in which 
$f_{crit}$ denotes the critical pressure fluctuation normalized by the unperturbed post-shock value. $M_{in}$ and $r_{sh}$ stand for the mass of the central compact object and the shock radius, respectively.
The critical fluctuation decreases with 
the shock radius, whereas it increases with the mass of the central object. We discuss the possible implications of our
results for 3D effects on shock revival, which is currently controversial in the supernova community.
\end{abstract}


\section{Introduction}
\label{sec:intro}
Decades of research on core-collapse supernovae (CCSNe) have not accomplished full understanding of the explosion mechanism yet. 
One of the most important issues in supernova theory is the dynamics of shock wave in the stellar core. In fact, it is widely accepted that the prompt shock wave, which is generated by core bounce, experiences 
stagnation at $r \sim 100$km owing to the energy losses by photodissociations of heavy elements and neutrino emissions as well as due to the 
ram pressure of accreting matter. For successful explosion, the stalled shock wave needs re-invigoration one way or another. 
The most promising is the heating by neutrinos diffusing out of a proto-neutron star. Only a percent of the total amount of 
energy neutrinos are carrying is sufficient to give the canonical explosion energy of supernova ($\sim 10^{51}$erg). Although it is still 
unclear whether the energy transfer from neutrinos to ejecta is indeed large enough, the neutrino heating is 
currently the most favored mechanism of CCSNe (see e.g. \citet{2012arXiv1204.2330K} for a recent review).

In the last five years, we have witnessed some successful explosions in advanced multi-dimensional numerical simulations.
(see e.g. \citet{2006ApJ...640..878B,2008ApJ...685.1069O,2009ApJ...694..664M,2010PASJ...62L..49S,2012ApJ...749...98T,2010arXiv1002.4914B,2012arXiv1202.0815M}. The results of different groups still have some discrepancies, though. In particular, 3D effects are still controversial. Although large-scale numerical simulations are 
making a rapid progress toward sufficient reality, we believe that phenomenological approaches such as proposed in this paper can 
still play an important and complimentary role for extracting key physical elements from complex and non-linear dynamics of CCSNe.

It was \citet{1993ApJ...416L..75B} who took the lead in such an approach. They examined a sequence of steady accretion flows 
through a standing shock wave onto a proto-neutron star to discuss the criterion for shock revival. They varied the luminosity of electron-type neutrinos ($L_{\nu_{e}}$) and mass accretion rate ($\dot{M}$) as free parameters and revealed that there is a 
critical luminosity for a given mass accretion rate, above which no steady shocked accretion flow
obtains. They argued that the critical luminosity marks the trigger point for shock revival. This approach was later extended to rotational
configurations by \citet{2005ApJ...623.1000Y} and linear stability analysis was also applied \citep{2007ApJ...656.1019Y}. The latter authors 
pointed out that the critical luminosity could be smaller if one takes into account hydrodynamical instabilities such as radial
overstabilization modes, which were also observed in dynamical simulations (see e.g. \citet{2006ApJ...641.1018O}) as well as non-radial
modes. Although the approach certainly has limitations in reproducing full complexity of explosion mechanism, the critical luminosity 
has become one of the most useful measures for shock revival and some new analyses have been published in recent years 
(see e.g. \citet{2012ApJ...746..106P,2012ApJ...749..142F,2012arXiv1202.2359K} and references therein).

The existence of the critical luminosity has been also demonstrated in multi-dimensional numerical simulations with simplified treatments
of neutrino transfer such as the gray or light bulb approximation 
\citep{2006ApJ...641.1018O,2008ApJ...678.1207I,2008ApJ...688.1159M,2010ApJ...720..694N,2011arXiv1108.4355H}. 
There is a wide consensus at present that multi-dimensionally dynamics such as the standing accretion shock instability (SASI) 
(see e.g. \citet{2003ApJ...584..971B,2007ApJ...654.1006F}) and neutrino-driven convection (see .e.g. 
\citet{2011ApJ...742...74M,2012arXiv1205.3491M} and references therein) reduce the critical luminosity. This is mainly because  
the advection time scale tends to be longer owing to turbulent motions, which then leads to longer heating and provides a favorable 
condition for shock revival. In addition, the instabilities push the shock wave outward and expand the gain region. Hence the exploration 
of multi-dimensional neutrino-heating mechanism is currently the hot topic in the main-stream research on CCSNe.

In particular, 3D dynamics is one of the central issues. \citet{2010ApJ...720..694N} found in their 2D and 3D experimental simulations
that the critical luminosity is monotonically reduced with increasing spatial dimensions, i.e., 3D dynamics provides the conditions 
that are most favorable for shock revival. On the other hand, \citet{2011arXiv1108.4355H} obtained in their similar computations 
the results that are at odds with those presented by \citet{2010ApJ...720..694N}: they found no significant difference in the 
critical luminosity between their 2D and 3D models. The reason of the difference is not clear at the moment, since they made different
simplifying assumptions and employed different numerical techniques.

In this paper, we address the condition for shock revival again by a simplified phenomenological approach. We do not discuss the 
critical luminosity, however. Instead we propose to introduce the third parameter, {\it fluctuations}, and discuss the condition for shock revival in terms of them. In so doing we employ a semi-dynamical approach instead of dynamical simulations to approximately describe 
shock motions that are induced by fluctuations. This is an extension of the previous works that employed only steady states. One of the 
drawbacks in the latter approach is that by definition it cannot handle the temporal evolution of shock wave and, consequently, 
cannot address what will happen to the shock wave after it restarts to move. The semi-dynamical 
approach can remove these problems and demonstrate that there is a threshold of fluctuation amplitudes for a given combination of 
neutrino luminosity and mass accretion rate, beyond which shock revival occurs and a continuous outward propagation of the 
revived shock wave follows, which we think is the new criterion for successful explosions. 

The paper is organized as follows. In Section 2, we describe the semi-dynamical model.
Confirming that it can capture the shock dynamics by a comparison with dynamical simulations, we apply the model systematically to various initial conditions that are appropriate for the shock-stagnation phase in Section 3. Based on these results
we give the critical fluctuation amplitudes for shock revival. Finally, we discuss the possible implications of the new criterion 
for the multi-dimensional neutrino heating mechanism in supernova theory and give conclusions in Section 4.

\section{Semi-dynamical Method}
\label{sec:semidynami}
 We are interested in the phase at several hundred milliseconds after bounce, in which the prompt shock wave is stalled and matter 
flows through an almost steady shock and accretes onto a proto-neutron star; neutrinos are emitted from the neutrino sphere and heat 
up the gain region; the location of quasi-steady stagnated shock wave is determined by the neutrino heating and 
the ram pressure of accreting matter; the onset of shock revival is determined not by the non-existence of steady accretion flows but by 
some sort of hydrodynamical instabilities, e.g., radial over-stabilizing oscillations in spherical symmetry and non-radial SASI in 
multi dimensions~\citep{2006ApJ...641.1018O,2012ApJ...749..142F}.  In this paper we take a standpoint that the onset of shock revival is determined by the neutrino 
luminosity and mass accretion rate as well as fluctuations by the instabilities. Our semi-dynamical approach deals with the transition from
the quasi-steady accretion to the re-expansion of shock wave as well as the ensuing outward propagation of shock wave in 1D. Before 
going into detail, we first describe the essence of this new approach.

The semi-dynamical model begins with an addition of perturbation to the steady accretion shock by hand. Then the shock wave starts to move. 
We follow the subsequent shock motions not by hydrodynamical simulations but by the integration of a simplified equation of motion, which 
is base on the local Riemann problem. By local we mean that our formulation considers only the neighborhood of shock wave. This enables us to
avoid simulations. This local approximation is found to be reasonable agreement with the results of full dynamical simulations near the shock 
wave (see Section~\ref{subsec:stage2}, \ref{subsec:sec2rangeofappli} and Appendix~\ref{ape:reliablocalappro}).

The model calculations are 
computationally very cheap, by virtue of which we can investigate long-term ($\gtrsim 1 {\rm s}$) evolutions of shock wave for a large number 
of models with different backgrounds so that we could obtain the critical fluctuation amplitudes for shock revival very efficiently. 
Another benefit for the semi-dynamical approach is that it can study finite fluctuation amplitudes, which is in sharp contrast to linear 
stability analysis, in which the stability is considered only for infinitesimal perturbations. We stress that even if the shock wave is 
linearly stable, large enough perturbations may trigger shock revival, which we will see is really the case in the later section.

\begin{figure*}
\vspace{15mm}
\epsscale{1.0}
\plotone{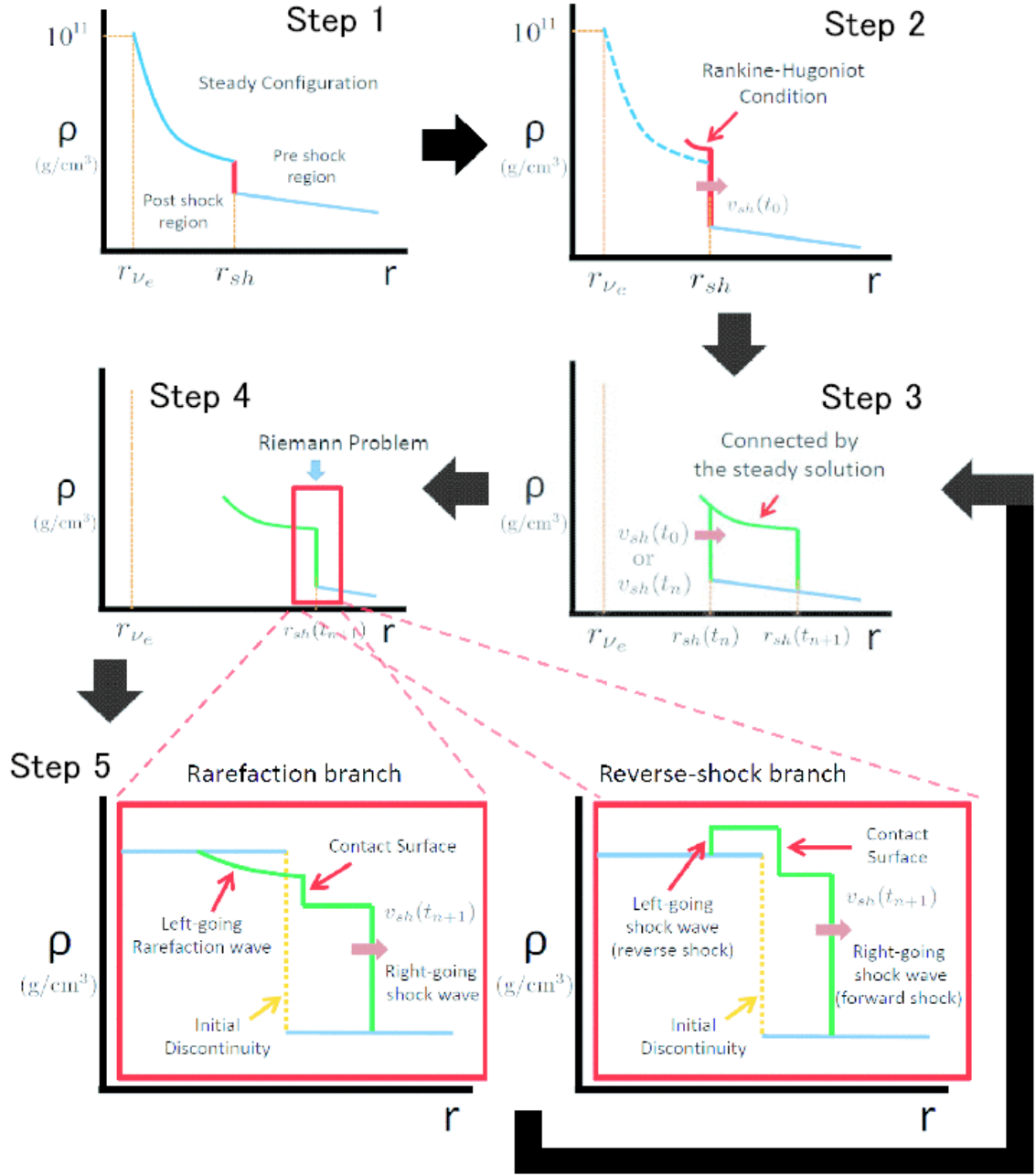}
\caption{Schematic pictures of individual steps in the semi-dynamical model. The horizontal axis in each panel denotes 
the radius whereas the vertical axis represents the density. See the text for more detailed explanations of each step.
\label{f1}}
\end{figure*}

\subsection{Details of the semi-dynamical model}
The actual calculations of shock motions in our semi-dynamical approach consist of a couple of steps, which we will describe more in detail in the 
following section (see also Figure~\ref{f1}).

\subsubsection{Step 1: preparation of steady shocked accretion flows}
\label{subsec:stage1}
The first step is a preparation of initial conditions, which are assumed to be steady and spherically symmetric. The procedure to obtain 
such flows is exactly the same as in \citet{2006ApJ...641.1018O} (see also \citet{2005ApJ...623.1000Y}) and the details are given 
in Appendix~\ref{ape:steadyshock}.

\subsubsection{Step 2: addition of perturbations}
\label{subsec:stage2}
After setting up the steady shocked accretion flows, we put some radial perturbations, which are not necessarily small, to the initial conditions. 
Although the actual form of perturbation is rather arbitrary, we choose in this study to give a finite velocity to the standing accretion shock 
wave by adjusting post-shock quantities so that the Rankine-Hugoniot relation would be satisfied for the given shock velocity, $v_{sh}(t_{0})$. 
In this expression, $t_{0}$ denotes the initial time.

\subsubsection{Step 3: displacement of shock wave}
\label{subsec:stage3}
The shock motions induced by the initial perturbations are calculated in a finite difference fashion. Suppose that the shock
location $r_{sh}(t_{n})$ and velocity $v_{sh}(t_{n})$ are given at $t_n$. Then the shock location at the next time $t_{n+1}$is
given by 
\begin{eqnarray}
&& r_{sh}(t_{n+1}) = r_{sh}(t_{n}) + v_{sh}(t_{n}) \times \Delta t \label{eqshevo},
\end{eqnarray}
where  $\Delta t = t_{n+1} -t_{n}$ is the interval between the two successive times. In this study we set $\Delta t = 10^{-4}{\rm s}$, 
which is sufficiently short. The initial time corresponds to $n=0$. If the shock velocity at $t_{n+1}$ is obtained somehow, then
the procedure is iterated until a designated time is reached.

\subsubsection{Step 4: determination of shock velocity -- setting up Riemann problem --}
\label{subsec:stage4}
How to obtain the new shock velocity $v_{sh}(t_{n+1})$ at $t_{n+1}$ is the most important part of this model. The idea here is 
similar to the one of the Godonov scheme in numerical hydrodynamics: a Riemann problem is set up at the beginning of each time step 
and the evolution during the subsequent interval is determined by the solution of the Riemann problem. In our model, we consider 
a certain Riemann problem at the new shock location $r_{sh}(t_{n+1})$ and solve it to find the velocity of the forward shock that 
exists always in the solution (see Step~5). 

In order to set up the Riemann problem, we need to specify hydrodynamical quantities on both sides of the discontinuity located 
at $r_{sh}(t_{n+1})$. The upstream quantities are easily obtained from the unperturbed steady accretion flow. The downstream quantities
are not so easy to obtain. In this model we assume that the downstream flow between $r_{sh} (t_{n})$ and $r_{sh} (t_{n+1})$ is steady.
We solve Eqs. (\ref{eqmasscon})-(\ref{eqelecfrac}) for steady flows from $r_{sh} (t_{n})$ to $r_{sh} (t_{n+1})$ to obtain the
hydrodynamical quantities at the latter point and use them for the Riemann problem. 

The post-shock flows are not steady. Since they are subsonic, it takes at least the advection time plus sound crossing time between the shock 
and the proto-neutron star to reach a steady state. In the shock revival phase, however, this time scale is longer than the typical 
time scales of variations in the shock radius, which means that the steady state is never realized over the entire post-shock region. 
On the other hand, we find in hydrodynamical simulations that the post-shock flows are approximately steady in the vicinity of the 
shock wave (see Appendix~\ref{ape:reliablocalappro}). By virtue of this approximation, we can avoid hydrodynamical simulations 
in the entire region and instead treat the shock motion alone. 

\subsubsection{Step 5: determination of shock velocity -- solving Riemann Problems --}
\label{subsec:stage5}
Now that the Riemann problem has been constructed, we can obtain the new shock velocity, based on the solution of the Riemann problem.
The bottom panels in Figure~\ref{f1} show the schematic pictures of two possible solutions of the Riemann problem. 
In the current situation, the solution always contain a forward shock wave, i.e., the right-going shock wave in Figure~\ref{f1},
which we regard as the displaced shock wave. The left-going wave can be either a rarefaction wave (left panel) or a shock wave (right
panel). Regardless, we adopt the velocity of the right-going shock wave as the new shock velocity at $t_{n+1}$, i.e, $v_{sh}(t_{n+1})$.
This closes a single time step. We go back to Step~3 and repeat the following steps for the next time step. The iteration is terminated
when the designated time is reached.

%\subsection{Miscellaneous}
\subsection{Miscellaneous}
\label{subsec:sec2rangeofappli}
In this study we fix the neutrino luminosity, mass accretion rate and mass of a central object during the time evolution 
for simplicity. It should be stressed, however, that the semi-dynamical model can handle the time dependence of these 
parameters with no difficulty. It is also important to note that the only difference between our model and full (1D) hydrodynamical 
simulations is how to obtain the hydrodynamical quantities just behind the shock wave. Hence the accuracy of our model depends 
entirely on the validity of the ``locally-steady'' approximation. As will be demonstrated in \S\ref{sec:demon} and 
Appendix~\ref{ape:reliablocalappro}, it looks reasonably good in general. Since the post-shock flows are subsonic, they are always affected 
by the physical conditions of inner regions in principle. This will be particularly so if the relaunched shock wave is stalled again.
On the contrary, if the shock continues to propagate outward briskly, we expect that our approximation will work 
reasonably well.

%$t$(s)
%$v_{sh}$(cm/s)
%\\
%$r_{sh}$(cm)

%$v_{sh}(t_{n+1})$

%$v_{sh}(t_0)$ \\
%\hspace{1mm} or \\
%$v_{sh}(t_{n})$

\section{Results}
\label{sec:demon}
In this section we apply the semi-dynamical method to shock revival in the post-bounce supernova cores. We first study the characteristics of 
solutions and observe the existence of the critical fluctuations. Then we investigate the dependence of the critical fluctuation for shock revival
on some key parameters, which is our main result in this paper.
% Finally, we show the reliability of local approximation by comparing with the result of full dynamical simulations.

\subsection{Characteristics of shock evolutions}
\label{subsec:validity}

\begin{figure*}
\vspace{15mm}
\epsscale{1.0}
\plotone{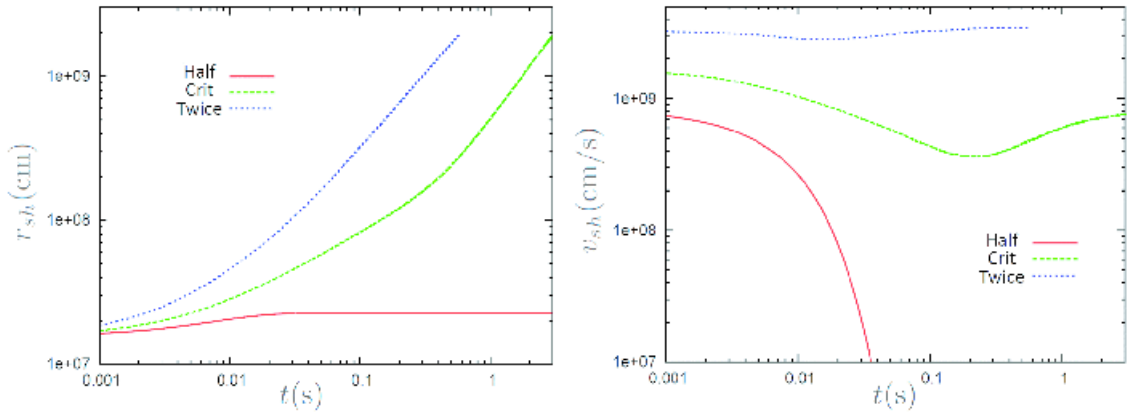}
\caption{The evolutions of shock wave for the fiducial background model. The left panel shows the shock radii as a function of time
whereas the right panel presents the corresponding shock velocities. Three lines with different colors in each panel correspond to 
different shock velocities added initially. See the text for more details.
\label{f2}}
\end{figure*}

We first investigate the evolutions of shock wave after the addition of perturbations of different amplitudes to the fiducial background 
model, which is characterized as follows: $L_{52} (\equiv L_{\nu}/({10^{52}}$erg/s))$=5$, $\dot{M}_{sun} (\equiv - \dot{M}/(1 M_{\odot}$/s))$=1$
% $L_{52} =5$, $\dot{M}_{sun} = 1$
 and $M_{in} = 1.4 M_{\odot}$ 
(see section~\ref{subsec:stage1})
% Here we define $L_{52} \equiv L_{\nu}/({10^{52}}$erg/s) and $\dot{M}_{sun} \equiv - \dot{M}/(1 M_{\odot}$/s) 
with $L_{\nu}$, $\dot{M}$ and $M_{in}$ denoting the neutrino luminosity, mass accretion rate and mass of a central object, respectively. 
In this fiducial model the standing shock wave is located at $r_{sh}=1.5 \times 10^{7}$cm.

Figure~\ref{f2} shows three representative evolutions of shock radius (left panel) and velocity (right panel). 
As demonstrated in the left panel, the shock wave either settles down to a new position (red line) or sustains the outward propagation (green and blue lines). 
The green line is in fact a dividing line of the two cases. These lines correspond to different initial shock velocities, with the blue and
green lines having the largest and smallest initial shock velocities, respectively. We refer to the initial perturbation for the green line as the critical shock velocity. 
%are divided into two cOne of them corresponds to the model with critical amplitude of perturbation for shock revival.
 The value of the critical 
shock velocity is $v_{sh(crit)} = 1.65 \times 10^{9}$cm/s in this case.

As is evident from the right panel, the out-going shock wave is decelerated initially in all cases. The model for the red line, which is given half the critical shock velocity initially, finds continuous deceleration of shock wave until it comes to a halt at $t \sim 40{\rm ms}$.
On the contrary, in the models of the green and blue lines, 
the latter of which is given twice the critical shock velocity at the beginning, the shock wave begins to accelerate at some point 
of time ($t \sim 0.2{\rm s}$ for the green line and $t \sim 0.01{\rm s}$ for the blue line) and keeps the outward motion up to the end of 
calculation. It is also clear that the shock wave evolves faster for larger initial shock velocities among the models that do not fizzle out.

The deceleration is efficient until the shock wave reaches $r_{sh} \sim 10^{8}{\rm cm}$ (see Figure~\ref{f2}). In fact, 
it seems that the shock waves that cannot make this distance fail to revive and vice versa. The shock deceleration in the early phase of
shock revival is consistent with the fact that the fiducial background model is linearly stable against radial perturbations. Note that the 
linear stability of spherically symmetric, shocked accretion flows can be judged by the Nakayama's criterion 
\citep{1994MNRAS.270..871N,1996MNRAS.281..226N}, which states that such flows are linearly stable if the post-shock 
matter is decelerated (see also \citet{2007ApJ...656.1019Y}). We confirm that this is indeed the case for the fiducial model employed here. 
Of the three models shown in Figure~\ref{f2}, the failed case (red line) has relatively small initial perturbations, 
and linear analysis will be applicable. To the other two cases, which are given larger perturbations, linear analysis may not be
applicable. We think, however, the cause of the shock deceleration can be understood in the same way also in these cases at least qualitatively. It is then nice that the results of semi-dynamical method are consistent with the linear stability analyses.

\begin{figure*}
\vspace{15mm}
\epsscale{1.0}
\plotone{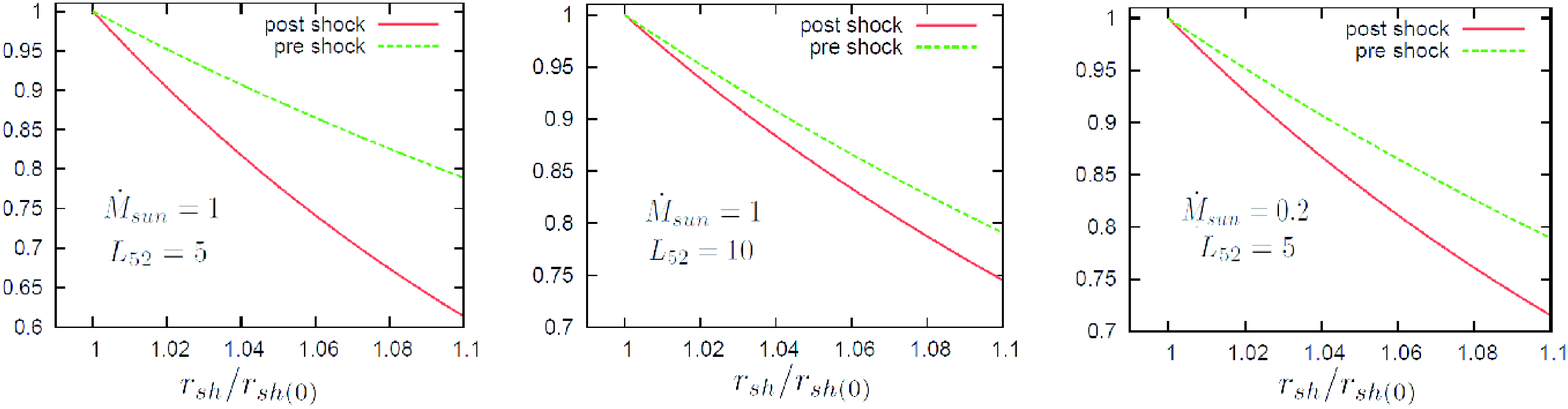}
\caption{Momentum fluxes just ahead of and behind the shock wave if it were moved to the position specified by the  
the horizontal axis. The shock radius is normalized by the original radius, $r_{sh(0)}$. The vertical axis denotes the 
momentum flux normalized by the value at $r_{sh(0)}$. The red solid lines give the post-shock values whereas the green dotted 
lines present the pre-shock ones. The left panel shows the momentum flux imbalance for the fiducial background model whereas 
the other two panels represent different background models with the neutrino luminosity and mass accretion rate displayed in 
each panel. Both models have larger shock radii initially than the fiducial model: $r_{sh(0)}=9.8 \times 10^{7 }{\rm cm}$ for the 
model in the middle panel whereas $r_{sh(0)}= 6.4 \times 10^{7}{\rm cm}$ for the model in the right panel.
\label{f3}}
\end{figure*}

The shock deceleration may be understood yet another way. \citet{1994PASJ...46..257N} claimed that the stability of 
standing shock wave can be judged by the momentum-flux imbalance between the pre- and post-shock flows. Of course they are exactly 
equal to each other for the initial standing shock wave. If the shock wave is shifted slightly outward from the original position, then the momentum flux loses its balance and the shock wave will be either pulled back or pushed further. The former occurs if the momentum flux in the 
pre-shock flow is larger than that in the post-shock flow and this implies the stability of shock wave. The latter case corresponds to 
instability, on the other hand. Note that this stability criterion is consistent with that of Nakayama's 
(see e.g. \citet{2008ApJ...689..391N,2009ApJ...696.2026N,2010ApJ...711..222N}). It is also expected that the greater the imbalance of 
the momentum fluxes is, the stronger the deceleration or acceleration will be. We apply this criterion to the shock deceleration at 
$r_{sh} \lesssim 10^{8}{\rm cm}$.

We show in Figure~\ref{f3} the momentum fluxes just ahead of and behind the shock wave if it were moved to
the specified position by the initial perturbation. The three panels correspond to different background 
models (cf. Figure~10 in \citet{1994PASJ...46..257N}). The horizontal axis denotes the radius normalized by the unperturbed shock radius
$r_{sh(0)}$, whereas the vertical axis expresses the momentum flux normalized by the value at $r_{sh(0)}$. The red (green) lines give 
the momentum fluxes just behind (ahead of) the shock wave. The post-shock momentum flux is 
calculated by extending the steady post-shock flow up to the perturbed shock front. On the other hand, the pre-shock momentum flux  
is obtained by assuming that the pre-shock flow is unaffected by the perturbation.

The left panel of Figure~\ref{f3} corresponds to the fiducial background model. As is clear from the figure, the exact 
momentum balance is satisfied at the original shock location whereas the pre-shock momentum flux overwhelms the post-shock momentum flux
as the shock wave is moved outward. In this case, as already mentioned, the shock wave will tend to be pulled back to the original 
position, i.e, the shock wave will experience deceleration. We also note that the strength of deceleration differs among the
background models. The middle and right panels of Figure~\ref{f3} correspond to the cases either with a higher neutrino 
luminosity or with a lower mass accretion rate than the fiducial case. In both cases, the unperturbed shock waves are located at
larger radii than in the fiducial model. As is evident in these panels, the momentum flux imbalance between the pre- and post-shock 
flows is smaller than in the fiducial model. This indicates that the pull-back force that decelerates the shock wave gets weaker as 
the original shock radius becomes larger, the fact which is  important in analyzing the critical fluctuation for shock revival in the next section.

\subsection{The Critical Fluctuation}
\label{subsec:critfluctu}

In the previous section, we show the existence of the critical shock velocity for the fiducial model, i.e., if the shock velocity administered
initially is larger than this value, the shock continues to propagate outward even if the neutrino luminosity is smaller than the critical
value. In this subsection, we perform a larger number of calculations for different background models, determining the critical 
shock velocity for each model, and then we analyze their property in detail.

%As shown below, the critical fluctuation depends strongly on the steady state configuration, especially the radius of standing shock location.

\begin{figure*}
\vspace{15mm}
\epsscale{1.0}
\plotone{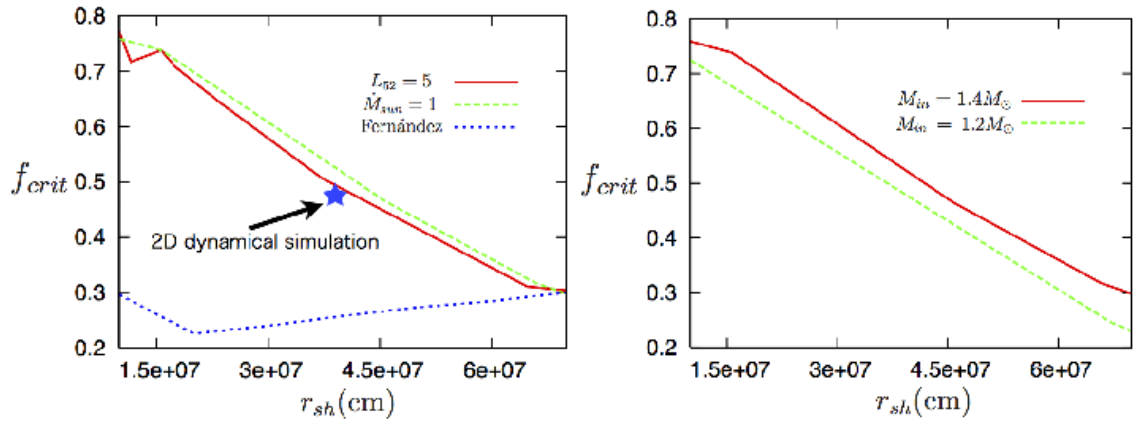}
\caption{The critical fluctuations as a function of the initial shock radius. The horizontal 
axis denotes the original shock radii in the unperturbed steady accretion flows, whereas the vertical axis represents the normalized 
critical fluctuations, which are defined by Eq.~(\ref{eqfluctu}). In the left panel the red line represents the sequence with a fixed 
neutrino luminosity ($L_{52}=5$) whereas the green line corresponds to the series with a fixed mass accretion rate ($\dot{M}_{sun} = 1$). 
The mass of the central object is set to $M_{in}=1.4 M_{\odot}$. The dotted line shows the pressure fluctuations that give the vanishing post-shock flow velocity. The star symbol displays the result of the 2D simulation detailed in Appendix~\ref{ape:dynamicalsimulation}.
 In the right panel, on the other hand, we compare the results for 
different masses of the central object. The red line represents the model with $M_{in}=1.4 M_{\odot}$ whereas the green 
line corresponds to the model with $M_{in}=1.2 M_{\odot}$. The mass accretion rate is fixed ($\dot{M}_{sun} = 1$) and 
the neutrino luminosity is varied in this case.
\label{f4}}
\end{figure*}

We prepare the unperturbed steady accretion flows as follows: we fix the mass of a central object to $M_{in} = 1.4 M_{\odot}$; in one 
sequence we vary mass accretion rate with a fixed neutrino luminosity (the red line in the left panel of Figure~\ref{f4}) 
whereas we fix the mass accretion rate and change the neutrino luminosity in another series (the green line in the same figure). 
Note that in these sequences of background models, the location of shock wave ($r_{sh}$) has one-to-one correspondence either with the mass accretion rate (for the red line) or with the neutrino luminosity (for the green line). The dependence on the mass of central object will be studied later.

Having in mind that the initial perturbations are generated by the hydrodynamical instabilities in reality, we refer to the 
dimensionless perturbation in the post-shock pressure, which is defined to be 
\begin{eqnarray}
&& f \equiv (p - p_{0})/p_{0} \label{eqfluctu},
\end{eqnarray}
as the fluctuation hereafter. In this expression $p$ and $p_{0}$ stand for the post-shock pressures for the perturbed and 
unperturbed flows, respectively. The value of the fluctuation that corresponds to the critical shock velocity is called the critical
fluctuation. Figure~\ref{f4} shows the critical fluctuations $f_{crit}$ as a function of the initial radius of 
shock wave ($r_{sh}$) for the two series of background models mentioned above. The critical fluctuation for each background model 
is obtained by many trials of semi-dynamical calculations with various initial shock velocities.

It is beyond the scope of this paper to address the nature of the fluctuations, e.g. kinematic or thermal. We infer, however, that the shock wave could be revived equally by the increase in temperature (or internal energy). It is noted that the increase in the post-shock temperature by $\sim 20 \%$ corresponds to the rise in the internal energy that we see for the critical shock velocity.

As we can see from the left panel of Figure~\ref{f4}, the critical fluctuation becomes smaller with the  
shock radius. It is interesting that the two lines are almost identical in the range of 
$10^{7}{\rm cm} \lesssim r \lesssim 6 \times 10^{7}{\rm cm}$. This is the most important region in discussing shock revival, since 
it roughly corresponds to the typical location of the stagnated shock wave. The above results suggest that the critical fluctuation
is mainly determined by the shock location and that the neutrino luminosity and mass accretion rate do not directly dictate shock 
revival; instead, they are indirectly important in the sense that they determine the initial location of standing shock wave. 
The reason why large radii of standing shock wave are advantageous for shock revival can be understood from the momentum-flux imbalance. 
As explained in \S~\ref{subsec:validity} and Figure~\ref{f3}, the pull-back force exerted on the shock wave   
is weakened with the radius of standing shock wave and hence smaller fluctuations are sufficient for shock revival at larger distances.
%\naga{It is also important to note that if we measure the critical fluctuation by the amplitude of temepature fluctuation ($f_{T} \equiv (T-T_{0})/T_{0}$), $f_{T}$ is $\sim 4$ times smaller than $f_{crit}$ (since the post-shock flow is radiation-dominated). It indicates that if the shock wave is unmoved but the temperature in the pos-shock region increase by $sim 10 \%$, the stagnated shock wave is expected to be revived.}

 It is interesting to note that \citet{2005ApJ...623.1000Y} observed that the shock radius at the critical point does not differ much among various models,
% almost the identical shock radius for various models at their critical points,
 which suggests that the shock location is the important quantity to determine the critical point. 
\citet{2008ApJ...688.1159M} conducted experimental simulations in multi-dimensions and found that the larger neutrino luminosity 
induces the greater amplitude of shock oscillations, which is also consistent with our interpretation, since the  high neutrino 
luminosities result in large shock radii and the pull back force acting on the shock wave is then weak.

We next vary the mass of central object to see the dependence of our findings on the change.
Fixing the PNS mass is admittedly inconsistent with the accretion flows with the substantial mass accretion rates adopted in this paper. 
The semi-dynamical method considers only the vicinity of shock wave. 
This means that we can not trace the evolution of PNS mass precisely, 
since the mass accretion rate ahead of the shock is different from that at PNS 
(note that the post-shock flow is no longer steady). 
Although, only the included mass is important in spherical symmetry and we can evaluate it by time-integrating the mass accretion rate, we do not think even this is necessary at the current level of approximation.
Instead we have chosen to change the assumed (constant) PNS mass and study the dependence of critical fluctuations on it. 
%We think that the study of shock revival in the dynamical background (time-dependent mass accretion rate, luminosity, neutrino temperature at neutrino sphere and PNS mass) is important and will be a future work based on hydrodynamical simulations. 

The results are shown in the right panel of Figure~\ref{f4}. This time we investigate the sequence of steady accretion flows, in which the mass accretion 
rate is fixed ($\dot{M}_{sun} = 1$) and the neutrino luminosity is varied. The result would not be changed if the sequence with the 
fixed neutrino luminosity were studied. The red line in the figure represents the critical fluctuations for $M_{in} = 1.4 M_{\odot}$ 
(identical to the green line in the left panel of Figure~\ref{f4}) whereas the green line denotes the critical fluctuations
for $M_{in} = 1.2 M_{\odot}$. Although the difference between the two cases is not so large, the critical fluctuation 
for $M_{in} = 1.4 M_{\odot}$ is systematically larger by several percentage points than that for $M_{in} = 1.2 M_{\odot}$ for the 
same initial shock radius. This is mainly attributed to the fact that gravity is stronger for the heavier central object, which then 
leads to the greater ram pressure. The result clearly demonstrates that larger masses of the central core are negative for shock revival, 
which is consistent with the analysis based on the critical luminosity by \citet{2012arXiv1202.2359K}.

% As a result, the ram pressure becomes weaker and the effect of neturino heating becomes more significant than larger mass of central core.

% These results are consistent with the dependence on $M_{in}$ in the steady state (smaller $M_{in}$ leads to have larger $r_{sh}$, which is favorable condition for the explosion as discussed in subsection~\ref{subsec:sphesteady}).

We find that the results given in Figure~\ref{f4} can be approximated by
\begin{eqnarray}
&& f_{crit} \sim 0.8 \times (\frac{M_{in}}{1.4M_{\odot}}) \times \{ 1 - (\frac{r_{sh}}{10^{8} {\rm cm}})  \}. \label{eqfluctucrit}
\end{eqnarray}
This simple analytic expression will be useful in analyzing the onset of explosion in full dynamical simulations, since 
the fluctuation $f$ can be easily estimated at each time step. Note that $p_{0}$ is obtained from the values of hydrodynamical quantities
just ahead of the stalled shock wave by using the Rankine-Hugoniot relation for $v_{sh}=0$. 
We expect that if the fluctuation $f$ at a certain time step in a simulation does not exceed $f_{crit}$ estimated this way, the shock wave 
does nothing but oscillations around the average position.

It is interesting to compare our results with those by \citet{2012ApJ...749..142F}, which propose a sufficient condition for shock revival in spherical symmetry. According to their analysis, the shock wave starts a runaway expansion when a portion of the fluid in the post-shock flow achieves positive energy. They also find that this happens when the post-shock velocity becomes positive. Motivated by these findings, we calculate for various background models the fluctuations that give the vanishing fluid velocity just behind the shock wave. If the unperturbed state is unstable to a radial overstabilizing mode and the criterion for shock revival by \citet{2012ApJ...749..142F} holds, the fluctuation obtained this way should be the true critical fluctuation. We find that they are smaller than by a factor of 2-3 the critical fluctuations obtained by the semi-dynamical method (See Figure~\ref{f4}). This difference may be attributed to the approximation employed in the semi-dynamical approach, i.e. the post-shock flows are assumed to be determined locally, which seems indeed to be a rather poor approximation in the shock revival by the overstabilization mode. In this sense, we may claim that our estimate of the critical fluctuation (Eq.~(\ref{eqfluctucrit})) is conservative.

It is worth noting that shock revival may be induced not by the overstabilization but by the multi-dimensional instabilities in reality. We confirm by 1D and 2D hydrodynamical simulations (see Appendix D) that the fiducial model in this paper is stable to the overstabilizing mode and does not produce shock revival in 1D, whereas large-amplitude fluctuations generated by SASI and/or neutrino driven convection revive the stalled shock wave in 2D. This is consistent with the previous papers
\citep{2006ApJ...641.1018O,2007ApJ...656.1019Y}. The 2D model is also used to demonstrate a possible application of the critical fluctuation to the analysis of multi-D shock revival. In Figure~\ref{f10}, we show the pressure distributions along a certain radial ray around the shock revival in the 2D simulation. It is interesting that the pressure fluctuation obtained from this figure is comparable to the critical fluctuation given by Eq.~(\ref{eqfluctucrit}) as shown in Figure~\ref{f4}.
Note, however, that this is just a single demonstration and the systematic comparison with full, 2D and 3D simulations 
is needed before we can make any quantitative assertion. Nonetheless the result certainly warrants further investigations.

It is also important to note that the critical fluctuation given by Eq.~(\ref{eqfluctucrit}) is a conservative estimate, since we ignore the time evolution of 
mass accretion rate in this analysis. Since the mass accretion rate becomes smaller with time in reality, the ram pressure at shock front,
which is one of the main obstacles for shock propagation, should be smaller, which would likely help the shock wave move outward. 
In fact, we confirm in another series of semi-dynamical calculations that decrease the mass accretion rate can induce shock revival
(see Appendix~\ref{ape:decreaseacrate}).

\section{Summary and Discussions}
\label{sec:discussion}

\begin{figure*}
\vspace{15mm}
\epsscale{1.0}
\plotone{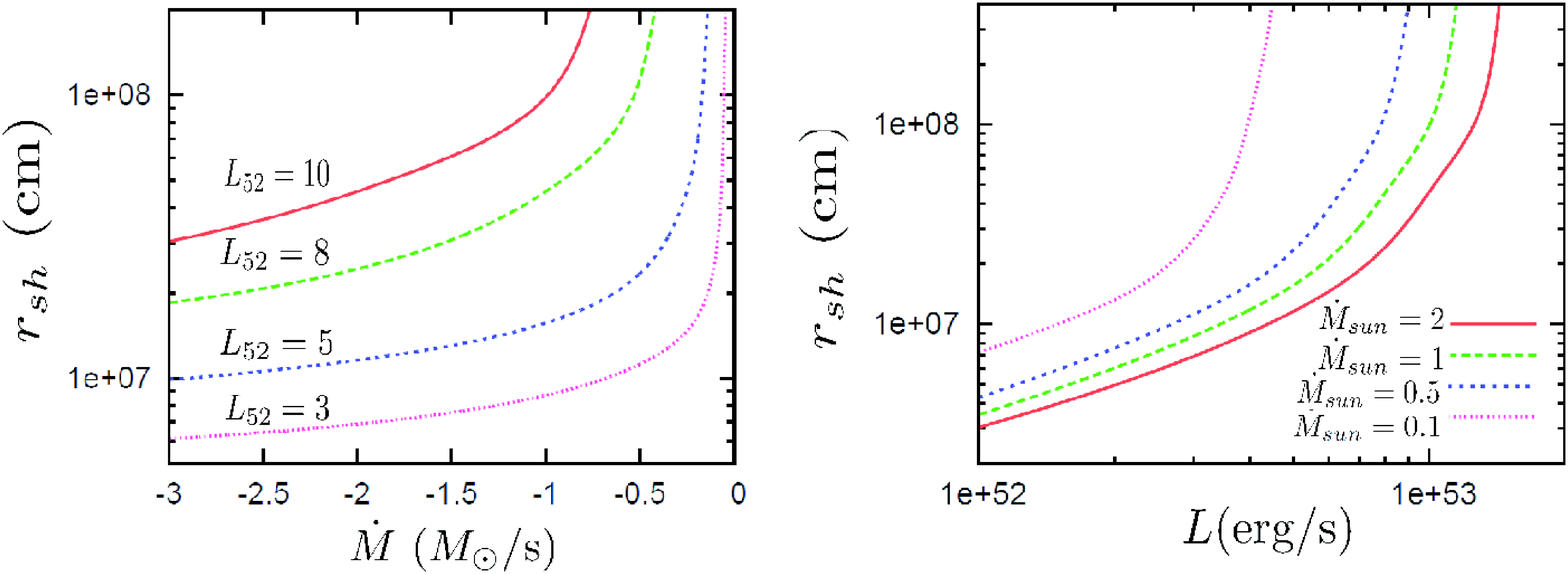}
\caption{Radii of standing shock waves in unperturbed, steady accretion flows. In the left panel, we vary the mass accretion rate for 
fixed neutrino luminosities whereas in the right panel, we fix the mass accretion rate and change the neutrino luminosity
for each line.
\label{f5}}
\end{figure*}

In this paper, we have developed a new approach to shock revival in the neutrino heating mechanism in core-collapse supernova.
The semi-dynamical model takes into account the temporal evolution of shock wave induced by 
fluctuations that are possibly generated by instabilities.
 The equation of motion of the shock wave is modeled in a finite difference manner, with Riemann problem being set up at 
each time step. In so doing the post-shock flows are approximated by locally steady flows, which are indeed observed in hydrodynamical
simulations. The method is very efficient in conducting a large number of model calculations of the long-term shock evolution.

As a result, we have demonstrated that there is a critical fluctuation for a given steady accretion flow, above which the shock 
can sustain outward propagation, leading hopefully to a successful supernova explosion. Varying the neutrino luminosity or mass accretion
rate for different masses of the central object, we have obtained a simple fitting formula for the critical fluctuation as a function of 
the initial shock radius and mass of central object (see Eq.~(\ref{eqfluctucrit})). According to our results, the critical fluctuation 
decreases with an increase in the shock radius and/or a decrease in the mass of central object. In particular, we have found that 
the initial shock radius is the key parameter for shock revival.

It should be noted that in our models the neutrino luminosities are
all sub-critical and there is a steady shocked accretion flow for a given pair of neutrino luminosity and mass accretion rate. 
In fact, even if the shock wave is stagnated around $r_{sh} \sim 10^{7}{\rm cm}$, in which case the neutrino luminosity is much 
smaller than the critical value, a large enough fluctuation can put the shock wave into a sustained outward propagation.
We have hence concluded that although the neutrino luminosity and mass accretion rate are important in determining the initial location of 
stagnated shock wave, it is the dynamical fluctuations that have direct leverage on shock revival.

Based on our findings in this paper, we now discuss possible implications for the neutrino heating mechanism in multi-dimensions.
 According to the recent results of 3D core-collapse simulations by \citet{2012ApJ...749...98T,2012arXiv1210.5241D}, the maximum residency 
time of accreting matter in the gain region is longer in 3D than in 2D owing to the extra degree of freedom in motion.
% This leads to longer neutrino heating, which then results in larger average radii of stagnated shock wave in 3D than in 2D (CHECK THIS!!!).
This leads to longer neutrino heating, which then results in larger average radii of stagnated shock wave in 3D than in 2D.
According to our findings in this paper, the larger shock radius means a smaller critical fluctuation for shock revival. 
In this sense at least 3D models are more favorable for successful explosions. However, the fluctuations generated by SASI or 
neutrino-driven convections are smaller in 3D than in 2D, since the free energy of turbulence could be distributed over a larger
number of oscillation modes in 3D \citep{2010ApJ...720..694N,2008ApJ...678.1207I} The inverse cascading nature of 2D turbulence may also contribute \citep{2012ApJ...759....5B}. In fact, the sloshing modes, which are always observed 
markedly in 2D axisymmetric simulations, are not so remarkable in 3D with non-axisymmetric modes being dominant. These two factors 
(the initial shock location and the amplitude of fluctuations in hydrodynamical instabilities) compete with each other and 
make it difficult to unravel the net effect of 3D hydrodynamics, the fact that may lead to the current controversy between different 
groups~\citep{2010ApJ...720..694N,2011arXiv1108.4355H,2012ApJ...749...98T}.

Although it is beyond the scope of this paper to settle the dispute on the 3D effect, we can add some more considerations that may be useful.
Figure~\ref{f5} shows the locations of standing shock wave in unperturbed steady accretion flows as a function of 
mass accretion rate (left panel) or neutrino luminosity (right panel). In the former the neutrino luminosity is fixed whereas the
mass accretion rate is fixed in the latter. As shown clearly for both cases in this figure, the location of standing shock varies 
very rapidly as the mass accretion rate (left panel) or neutrino luminosity (right panel) exceeds a certain value. For instance, 
the blue line in the left panel, for which the neutrino luminosity is fixed at $L_{52}=5$, shows that $r_{sh}$ is almost constant
as long as $|\dot{M}_{sun}| \gtrsim 1$ whereas it grows rapidly for $|\dot{M}_{sun}| \lesssim 0.5$. A similar trend is evident in the 
right panel if one replaces the mass accretion rate with the neutrino luminosity. This may imply then that if the average shock radius 
is already in this rapidly changing regime, the shock radius of the two competing factors will be more important and 3D 
may be more advantageous for shock revival. If the shock wave is stagnated at rather small radii, on the other hand, the larger 
fluctuations in 2D will be more important to provide favorable conditions for shock revival.

In spite of its simplicity the semi-dynamical approach we have employed in this paper can capture the essential features of the shock 
dynamics such as linear stability. All the results we have obtained are of qualitative nature and the critical fluctuations will be
somewhat changed if more realistic treatments of hydrodynamics and microphysics are incorporated.
 We believe, however, that the existence
of the critical fluctuation and its qualitative dependence on the shock radius and mass of central object will not be changed qualitatively. 
It is true that there are some limitations to the semi-dynamical approach in this paper. In addition to the fact that it is 1D, 
some features in shock dynamics, such as re-stagnations, certainly cannot be described, since the locally-steady
approximation will not be valid. More detailed comparison with multi-D hydrodynamical simulations will reveal both the merit and demerit
of the approach, which will be the issue of our forth coming paper.

\acknowledgements 
This work was supported by Grant-in-Aid for the Scientific Research from the Ministry of Education, Culture, Sports, Science and Technology (MEXT), Japan (21540281, 24244036, 22540297, 24740165) and HPCI Strategic Program of Japanese MEXT.

\appendix
%\section{Appendix A: Construction of Steady Shocked Accretion Flows}
\section{Construction of Steady Shocked Accretion Flows}
\label{ape:steadyshock}
In this section, we present details in the construction of unperturbed, spherically symmetric, steady accretion flows in Step~1 
of \S\ref{subsec:stage1}. We solve the following ordinary differential equations from the shock front ($r_{sh}$) to the neutrino 
sphere for electron-type neutrinos ($r_{\nu_{e}}$): 
\begin{eqnarray}
&& 4 \pi r^2 \rho v = \dot{M} \label{eqmasscon}, \\
&& v \frac{dv}{dr} + \frac{1}{\rho} \frac{dp}{dr} + \frac{GM_{in}}{r^{2}} = 0 \label{eqmomentumcon}, \\
&& v \frac{d}{dr} (\frac{e}{\rho}) - \frac{p}{ \rho^{2}} v \frac{d \rho}{dr} = \frac{Q_{E}}{\rho} \label{eqenecon}, \\
&& v \frac{dY_{e}}{dr} = Q_{N} \label{eqelecfrac}.
\end{eqnarray}
In the above expressions, $\rho$, $p$, $T$, $e$ and $Y_{e}$ denote the mass density, pressure, temperature, energy density 
and electron fraction, respectively. Other symbols, $r$, $v$, $G$, $\dot{M}$, $M_{in}$, stand for the radius, fluid velocity, 
gravitational constant, mass accretion rate and mass of a central object, respectively. Here Newtonian gravity is assumed and 
the self-gravity of accreting matter is neglected. Interactions between neutrinos and matter are encapsulated in $Q_{E}$ and $Q_{N}$,
the expressions of which are adopted from Eqs. (16) and (17) in \citet{2006ApJ...641.1018O}. We employ the equation of state 
(EOS) by \citet{2011ApJS..197...20S}, which is based on the relativistic mean field theory.
% For all the models investigated in this paper, the mass of the central object and electron (anti-electron) type neutrino temperature are fixed to $M_{in} = 1.4 M_{\odot}$, $T_{\nu_{e}(\bar{\nu_{e}})} = 4 (5)$MeV, the typical values in the post phase.

At the shock wave we impose the Rankine-Hugoniot relation for $v_{sh}=0$. The pre-shock flow is assumed to be a free fall
with the entropy per baryon of $s~=~3 k_{B}$ and electron fraction of $Y_{e}~=~0.5$. Here $k_{B}$ is the Boltzmann constant. 
At the neutrino sphere ($r_{\nu}$) the following relation is assumed:
\begin{eqnarray}
&& L_{\nu} = \frac{7}{16} \sigma T_{\nu}^{4} 4 \pi r_{\nu}^{2} \label{eqneutlumi},
\end{eqnarray}
where $L_{\nu}$, $\sigma$ and $T_{\nu}$ denote the neutrino luminosity, Stefan-Boltzmann constant and neutrino temperature, respectively. 
We further assume that the neutrino luminosities for electron-type neutrinos ($L_{\nu_{e}}$) and anti-neutrinos ($L_{\bar{\nu_{e}}}$) 
are identical (and denoted as $L_{\nu}$). We ignore other types of neutrinos, which play only a minor role in matter heating. 
For all the models investigated in this paper, the neutrino temperatures are fixed to $T_{\nu_{e}} = 4 {\rm MeV}$ and 
$T_{\bar{\nu_{e}}} = 5 {\rm MeV}$, the typical values in the post-bounce phase. Note that the neutrino spheres 
($r_{\nu_{e}}$ and $r_{\bar{\nu_{e}}}$) are uniquely determined for given luminosities in this prescription.

Once the shock radius ($r_{sh}$), mass of the central object ($M_{in}$), neutrino luminosity ($L_{\nu}$) and mass accretion 
rate ($\dot{M}$) are specified, a steady post-shock accretion flow is uniquely obtained as a solution of 
Eqs. (\ref{eqmasscon})-(\ref{eqelecfrac}). Imposing further that the density be $\rho = 10^{11}{\rm g/cm}^3$ at the inner boundary, which is assumed to coincide with $r_{\nu_{e}}$,
we determine the shock radius $r_{sh}$ Incidentally, we only consider the so-called inner solutions \citep{2005ApJ...623.1000Y,2012ApJ...746..106P}. The results given in Figure~\ref{f5} are also obtained this way. We also try a different inner boundary condition, i.e. the neutrino sphere be located at the radius, where the optical depth $\tau$ becomes 2/3, and find that it does not change the structure of accretion flow very much. Indeed, for the fiducial model, the radius of the unperturbed shock wave is shifted from $1.5 \times 10^{7}$cm to $1.2 \times 10^{7}$cm. This $ \sim 20 \%$ difference does not change the conclusion of this paper.

% In addition, we confirm that the different inner boundary condition, e.g. the neutrino sphere locates at the point where the optical depth from the shock becomes 2/3,

%pressure (erg/cm$^3$) \hspace{3mm}
%density (g/cm$^3$) \hspace{3mm} radius (cm)

%\section{Appendix B: Validity of the local approximation}
\section{Validity of the locally-steady approximation}
\label{ape:reliablocalappro}

\begin{figure*}
\vspace{15mm}
\epsscale{1.0}
\plotone{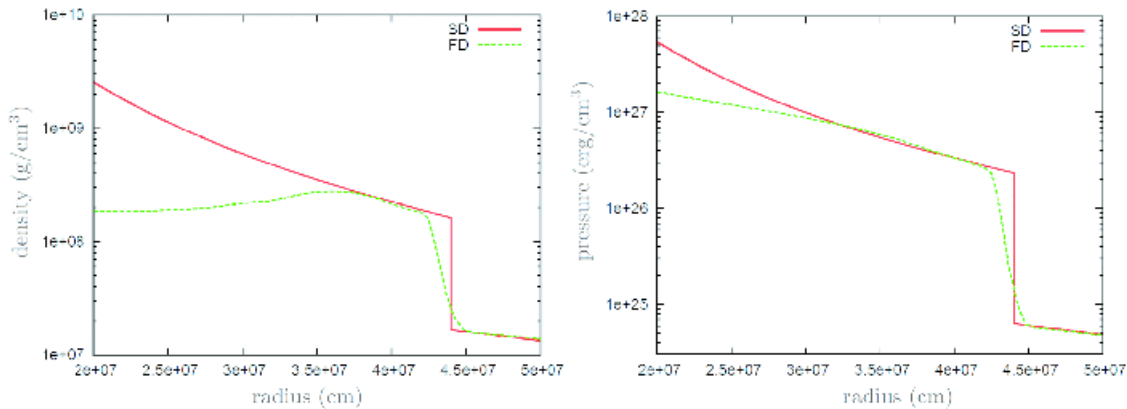}
\caption{Comparison of the radial profiles of density (left panel) and pressure (right panel) between 
a hydrodynamical simulation (green lines) and the corresponding globally steady accretion flow (red lines). See the text for details.
\label{f6}}
\end{figure*}

The semi-dynamical model assumes that the post-shock flows are locally steady near the shock wave.
As mentioned in the text, the reliability of the approach depends critically on the validity of this assumption.  Here we show a comparison with a result of hydrodynamical simulations by \citet{2012arXiv1209.4824Y}, which may support our assumption. The hydrodynamical models employ 
the so-called light-bulb approximation. We pick up a model of their computations, which has the following 
parameters: $L_{52}=5$, $\dot{M}_{sun} = 0.53$ and $M_{in} = 1.4 M_{\odot}$. We adopt the 
same values for these parameters also in the semi-dynamical model in this comparison. 

We take a snap shot from the hydrodynamical data and compare it with the corresponding steady 
post-shock configuration. For that purpose we need to extract the shock velocity in the 
hydrodynamical simulation.  We first identify the pre- and post- shock states and then calculate 
the shock velocity from the momentum flux balance. Although the numerical diffusion in dynamical simulations makes it difficult to identify the pre- and post-shock states very accurately, we decide to
adopt the hydrodynamical quantities at the points, where the density gradient starts to change rapidly. 
Although this is a rather crude treatment, it is sufficient for the present purpose. 

Figure~\ref{f6} shows the results of comparison. The left panel in this figure 
displays the density distributions whereas the right panel gives the pressure distributions. The red lines 
labeled as SD represent the globally steady accretion flows, whereas the green lines with the label FD give the results of 
the dynamical simulation. The shock velocity calculated from the momentum balance is 
$\sim 1.3 \times 10^{9}$cm/s in this case, which exceeds the critical shock velocity obtained 
by the semi-dynamical approach and, in fact, leads to shock revival in the dynamical simulation. 
As shown in this figure, the density and pressure distributions near the shock front are almost 
identical with those of the steady accretion flow. We hence confirm at least in this case that 
the approximation of local steadiness is good.

%\section{Appendix C: Shock evolutions induced by the reduction of mass accretion rate}
\section{Shock evolutions induced by the reduction of mass accretion rate}
\label{ape:decreaseacrate}
Although we fix the mass accretion rate in this paper for simplicity, it is a function of time in reality and
is an important factor to induce motions of stagnated shock waves. In this appendix, we consider
the shock evolution induced by the reduction of mass accretion rate imposed initially by hand instead
of taking fully into account the time-dependent mass accretion rate.

 After setting up a steady accretion flow (Step~1), we artificially reduce the mass accretion rate in the
 pre-shock region by a certain fraction. This naturally leads to a Riemann problem at the standing 
shock wave and induces its motion without additional perturbations to the shock velocity. The subsequent 
steps in the semi-dynamical method are completely the same as before (Section~\ref{sec:semidynami}). 
Note that the change of $\dot{M}$ is given only at the beginning and it remains constant 
afterwards. It is also assumed that the neutrino luminosity and mass of a central object
($M_{in} = 1.4 M_{\odot}$) are constant in this study.

\begin{figure*}
\vspace{15mm}
\epsscale{1.0}
\plotone{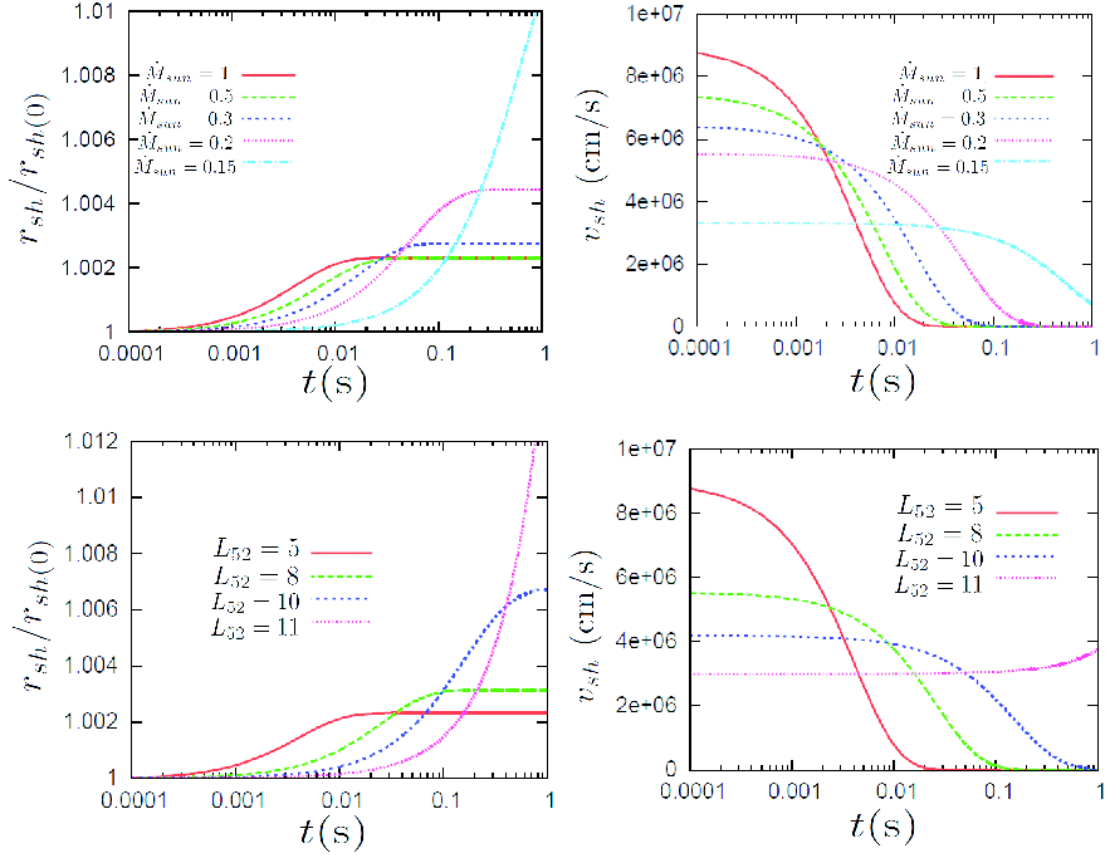}
\caption{The time evolutions of shock radius (left) and velocity (right) following the artificial reduction of the mass accretion 
rate by $1 \%$. In the upper panels, the background models have the same neutrino luminosity 
($L_{52} = 5$) but different mass accretion rates, whereas the background models for the 
lower panels have the identical mass accretion rate ($\dot{M}_{sun} = 1$) but 
different neutrino luminosities. 
\label{f7}}
\end{figure*}

%%%%%

\begin{figure*}
\vspace{15mm}
\epsscale{1.0}
\plotone{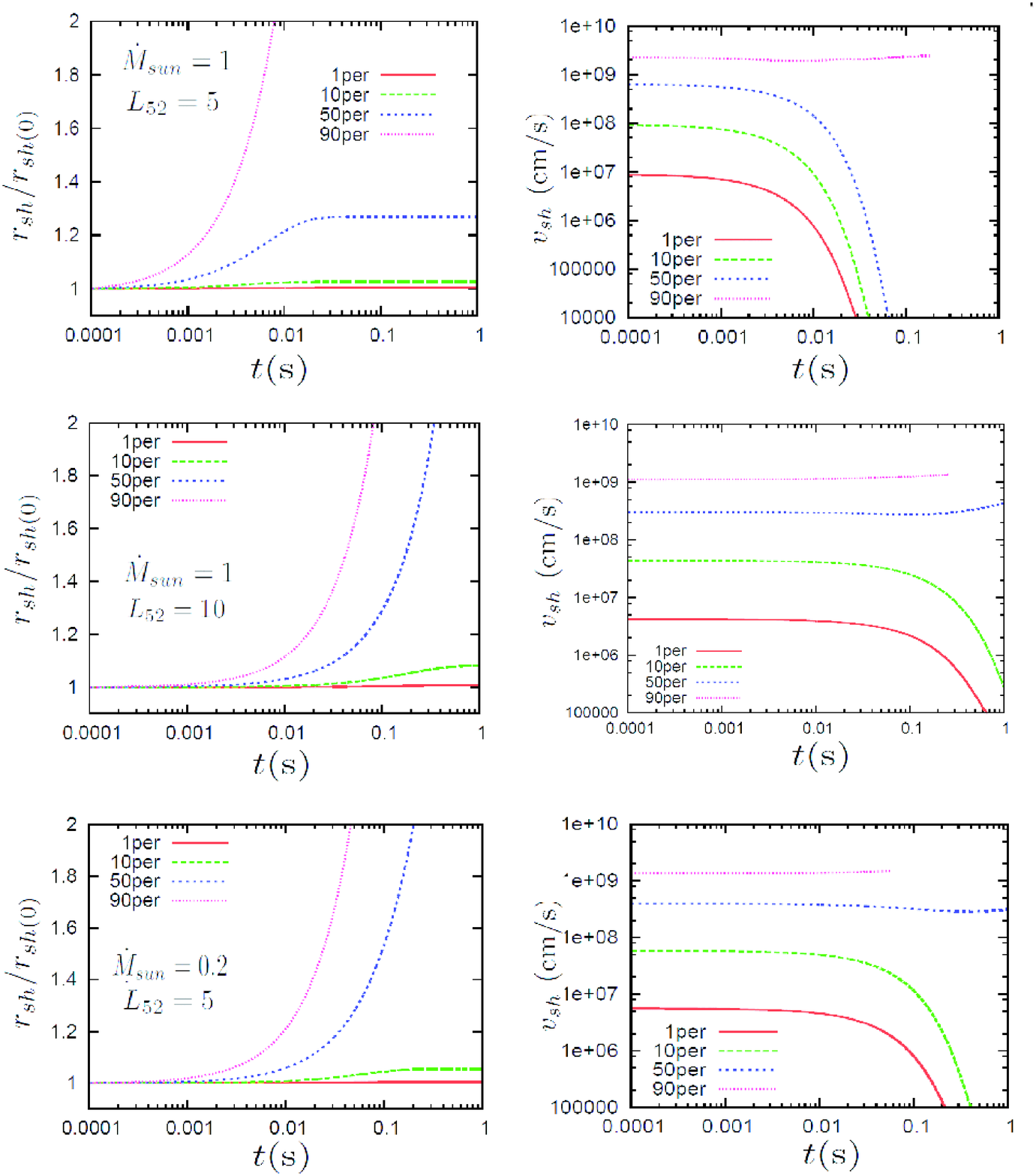}
\caption{The time evolutions of shock radius (left) and velocity (right) following the artificial reductions of the mass accretion 
rate by different fractions. The neutrino luminosities and mass accretion rates in the background 
models are given in the panels. The red, green, blue and pink lines correspond to the initial reductions of mass accretion rate by $1$, $10$, $50$ and $90 \%$, respectively.
\label{f8}}
\end{figure*}

%%%%

Figure~\ref{f7} shows the evolution of shock radius (left panels) and velocity (right panels)
for the models, in which $\dot{M}$ is reduced by $1 \%$ initially. The vertical axis of left panels 
is the shock radius normalized by the initial value, whereas the horizontal axis denotes the time 
after the onset of calculation. The upper panels present the results
for the background models with a fixed neutrino luminosity of $L_{\nu} = 5 \times 10^{52}$erg/s, whereas the lower panels display the results for the background models with a fixed mass accretion rate of $\dot{M}_{sun} = 1$.

In all models, the reduction of mass accretion rate in the pre-shock flow leads to the outward motions 
of the shock wave just as expected from the initial Riemann problems. Intuitively speaking, 
smaller mass accretion rates imply weaker ram pressures. As the neutrino luminosity increases or 
the mass accretion rate decreases in the background model and, as a consequence, the initial shock radius becomes larger, the shock reaches greater distances later. This trend is similar to the 
case considered in the main body, in which velocities are added to the standing shock wave initially. 
This is understood from the pull-back forces exerted on the shock wave just in the same way.

Next we change the initial reduction rate in the mass accretion rate of the pre-shock flows. Figure~\ref{f8} shows the evolutions of shock wave for 3 different background models. 
The upper panels correspond to the fiducial background model, whereas the middle (lower) panels 
show the results for the background model with $\dot{M}_{sun} = 1$ 
and $L_{52} = 10$ ($\dot{M}_{sun} = 0.2$ and $L_{52} = 5$). The left panels present the evolutions of 
shock radius and the right panels display the temporal changes of shock velocity. The red, green, 
blue and pink lines correspond to $1 \%$, $10 \%$, $50 \%$ and $90 \%$ reductions in the 
mass accretion rate, respectively. As is expected again, larger changes in the mass accretion rate 
push the shock wave outward more strongly (see Figure~\ref{f8}). In fact, 
we can consider the critical reduction rate in the mass accretion rate, below which the shock wave 
continues to move outward without re-stagnation. 

The critical reduction of mass accretion rate may be useful when one judges whether a hit of 
an interface between different layers, e.g., ${\rm Si/O}$ interface, on the stagnated shock wave
will lead to shock revival. It is also clear that a secular decline of mass accretion rate, which is
ignored in this paper, will be helpful for the shock propagation induced by the fluctuations in
hydrodynamical instabilities. In this sense the critical fluctuations given by Eq.~(\ref{eqfluctucrit})
is a conservative estimate.

%We find out that $ \sim 50 \%$ reduction of mass accretion rate is large enough to revive shock wave for lower two models in Figure~\ref{f8}.

\section{Hydrodynamical Simulations with Light Bulb Approximation}
\label{ape:dynamicalsimulation}

\begin{figure*}
\vspace{15mm}
\epsscale{1.0}
\plotone{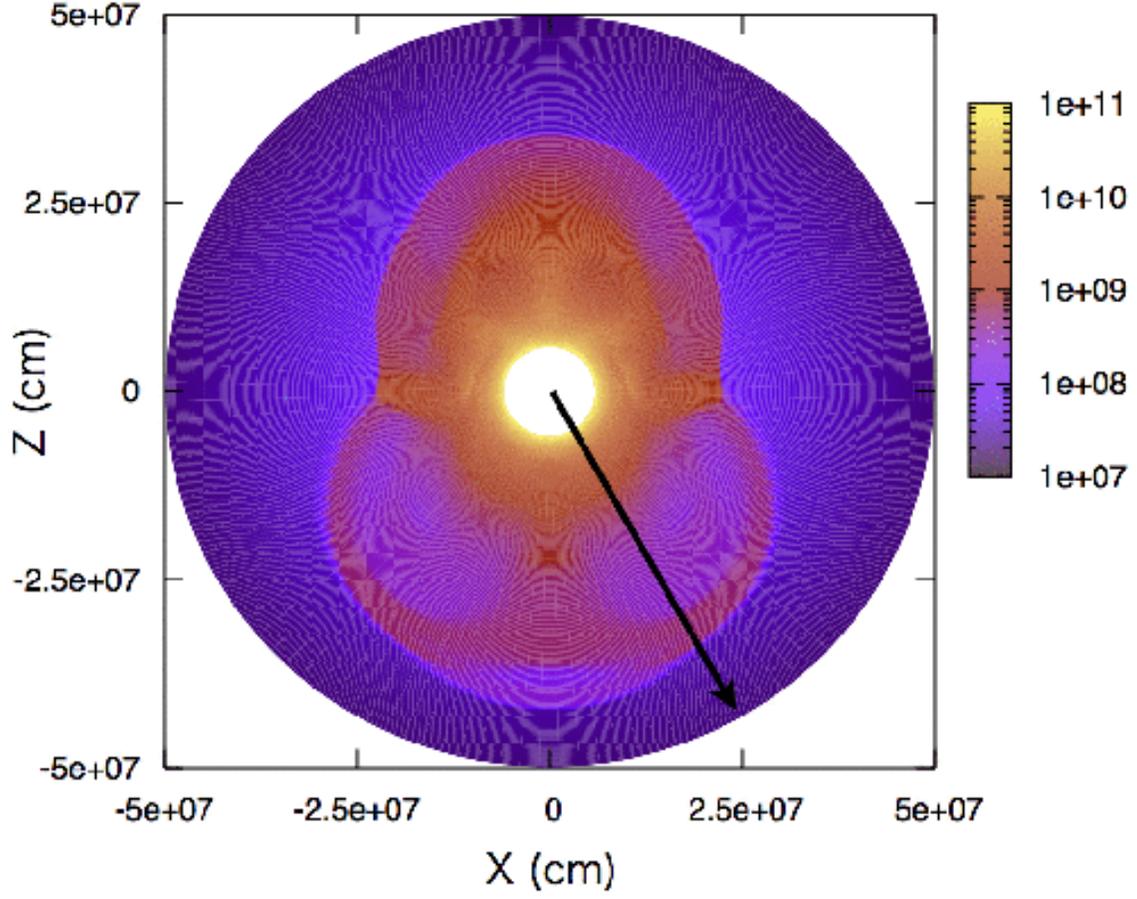}
\caption{The density distribution in the meridian section at $t=680$ms for the 2D dynamical simulation described in Appendix~\ref{ape:dynamicalsimulation}. The arrow represents the radial ray with $\theta = 150^{\circ}$, in which direction shock revival occurs. See also Figure~\ref{f10} for the temporal evolution of the pressure distributions along this line.
\label{f9}}
\end{figure*}

%%%%
\begin{figure*}
\vspace{15mm}
\epsscale{1.0}
\plotone{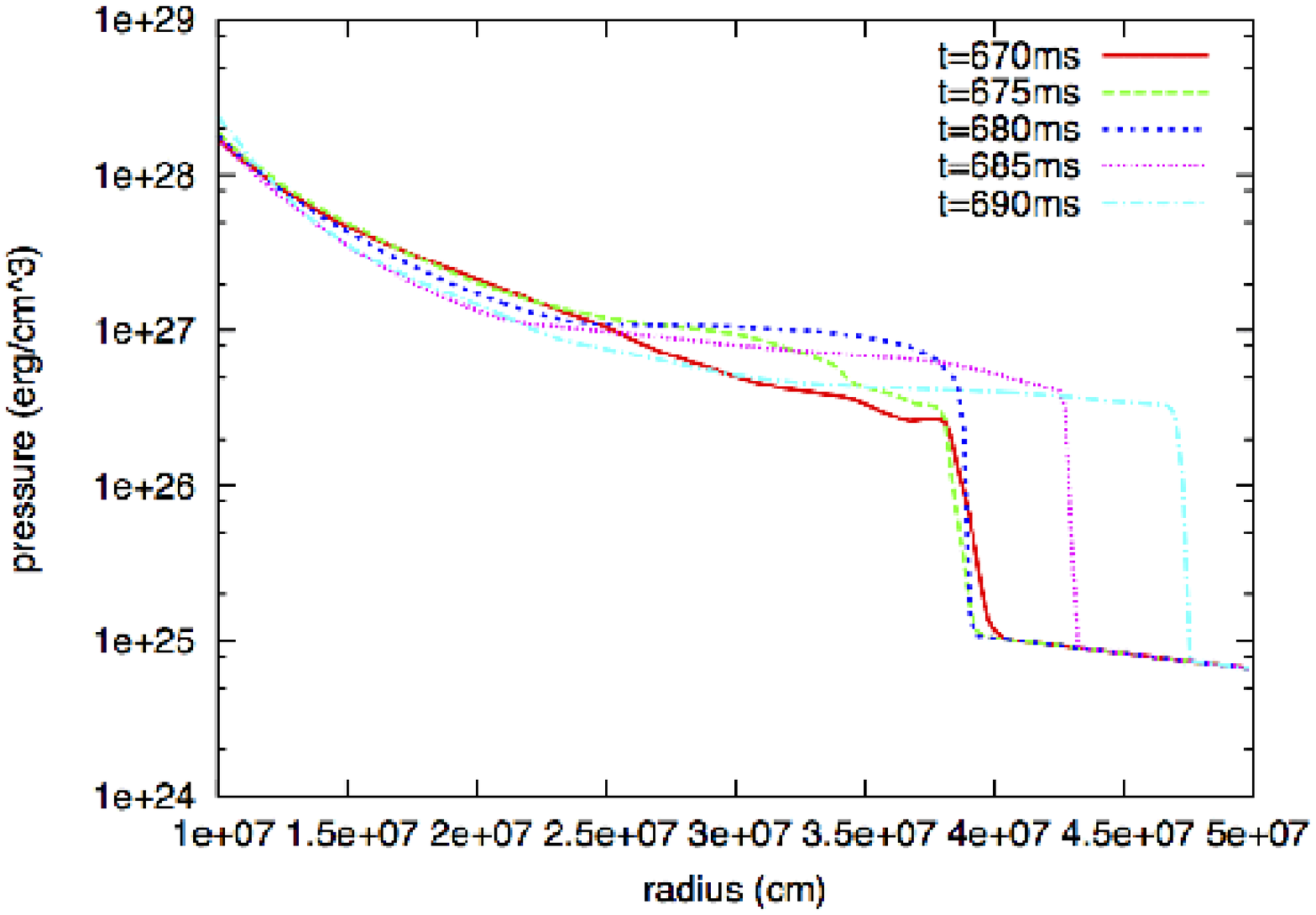}
\caption{The temporal evolution of the pressure distributions along the radial ray with $\theta=150^{\circ}$ around the time of shock revival. See also Figure~\ref{f9}.
\label{f10}}
\end{figure*}

In this paper, a small number of 1D and 2D hydrodynamical simulations are done to see the growth of hydrodynamical instabilities both radial and non-radial.
% both radial and non-radial
 The numerical code is based on the central scheme, which is a popular choice at present (see, e.g., \citet{2008ApJ...689..391N,2011ApJ...731...80N}). In this study, we employ the light bulb approximation and solve the following system of equations:
\begin{eqnarray}
&&\partial_{t} \biggl\{ r^2 \sin{\theta} \rho \biggr\} + \partial_{r}\biggl\{{ r^2 \sin{\theta} \rho v^r}\biggr\} + \partial_{\theta}\biggl\{{ r^2 \sin{\theta} \rho v^{\theta}}\biggr\} = 0,  \label{eq:hydi}\\
&&\partial_{t} \biggl\{ r^2 \sin{\theta} \rho v_{r} \biggr\} + \partial_{r}\biggl\{ r^2 \sin{\theta} \hspace{1mm} (\rho v^r v_r + p) \biggr\}
+ \partial_{\theta}\biggl\{ r^2 \sin{\theta} \hspace{1mm} \rho v^{\theta} v_{r} \biggr\} \nonumber  \\
&& = r^2 \sin{\theta} \biggl\{ - \rho \hspace{1mm} \frac{GM_{in}}{r^2} + r \rho (v^{\theta})^2  + \frac{2 p}{r} \biggr\},  \\
&&\partial_{t} \biggl\{ r^2 \sin{\theta} \rho v_{\theta} \biggr\} + \partial_{r}\biggl\{ r^2 \sin{\theta} \hspace{1mm} \rho v^r v_{\theta} \biggr\}
+ \partial_{\theta}\biggl\{ r^2 \sin{\theta} \hspace{1mm} (\rho v^{\theta} v_{\theta} + p ) \biggr\} \nonumber  \\
&& = r^2 \sin{\theta} \cos{\theta} \hspace{1mm} p,\\
&&\partial_{t} \biggl\{ r^2 \sin{\theta} ( e + \frac{1}{2} \rho v^2) \biggr\}  + \partial_{r}\biggl\{ r^2 \sin{\theta} v^r ( e+p + \frac{1}{2} \rho v^2) \biggr\} +
 \partial_{\theta}\biggl\{ r^2 \sin{\theta} v^{\theta} ( e+p + \frac{1}{2} \rho v^2) \biggr\} \nonumber  \\
&& =  - \sin{\theta} \hspace{1mm} \rho v^{r} {GM_{in}} + r^2 \sin{\theta} \hspace{1mm} Q_{E} , \\
&&\partial_{t} \biggl\{ r^2 \sin{\theta} \rho Y_{e} \biggr\} + \partial_{r}\biggl\{{r^2 \sin{\theta} \rho Y_{e} v^r}\biggr\}
+ \partial_{\theta}\biggl\{{r^2 \sin{\theta} \rho Y_{e} v^{\theta}}\biggr\} \nonumber  \\
&& = r^2 \sin{\theta} \hspace{1mm} \rho Q_{N},  \label{eq:hydf}
\end{eqnarray}
where $v^{r}(=v_{r})$, $v^{\theta}(=v_{\theta}/r^2)$ are radial- and angular-velocity in the meridian section and other notations are the same as Eqs. (\ref{eqmasscon})-(\ref{eqelecfrac}). 
%$T^{ij}$ is the energy momentum tensor of a perfect fluid, which is defined as:
%\begin{eqnarray}
%&&T^{ij} = \rho v^{i} v^{j} + p g^{ij},
%\end{eqnarray}
%where $g^{ij}$ is the space metric in spherical coordinates.
We assume no rotation and do not solve the azimuthal velocity. Note that the above equations are reduced to Eqs. (\ref{eqmasscon})-(\ref{eqelecfrac}) if the flow is steady and spherically symmetric.

The computational domain ranges from $r=r_{\nu_{e}}$ to $r=500$km and from $\theta = 0^{\circ}$ to $\theta = 180^{\circ}$ on the spherical coordinates. Note that no equatorial symmetry is assumed. We employ 300(r)$\times$ 60($\theta$) grid points. The radial grid is non-uniform, with the grid width being smallest ($\Delta r = 0.6$km) at the inner boundary and increased by 0.532$\%$ geometrically toward the outer boundary. We add an $l=1$ perturbation to the radial velocity at the initial step, where $l$ stands for the indices of Legendre polynomials. The amplitude of perturbation is chosen to be $1\%$.

We show the result for the fiducial model in this Appendix. It is noted first that this model gives no shock revival in the spherical symmetric computation, i.e. no overstabilization mode grows in this case. On the contrary the 2D axisymmetric simulation produces shock revival at $t \sim 700$ms. Figure~\ref{f9} shows the density map in the meridian section at $t=680$ms, just prior to the shock revived. Note that the average shock radius is increased by the non-radial instabilities. In fact, a strong $l=1$ sloshing mode develops during the evolution and leads finally to the shock revived on the South side.

 Figure~\ref{f10} shows the temporal evolution of the pressure distributions along the radial ray with $\theta = 150^{\circ}$ around the time of shock revival. The ray is depicted as an arrow in Figure~\ref{f9}. The shock wave is almost stagnated from $t=670$ms to $t=680$ms. It is also observed that a pressure wave is approaching the shock wave from inside and hits it at $t=680$ms. Then the shock wave starts to move outward and the shock revival obtains. Assuming that the pressure wave gives the critical fluctuation for shock revival at this point, we calculate the normalized fluctuation according to as Eq.~(\ref{eqfluctu}). The result, $f \sim 0.48$, is displayed as a star in Figure~\ref{f4}. Interestingly, the fluctuation obtained this way is very close to the theoretical estimate given by Eq.~(\ref{eqfluctucrit}), $f_{crit} \sim 0.56$. 

% The $f_{crit}$ derived from simple analytic expression in Eq.~(\ref{eqfluctucrit}) is around 0.56 for $r=4 \times 10^{8}$cm and $M_{in}=1.4M_{\odot}$. Therefore, the critical fluctuation derived from Eq.~(\ref{eqfluctucrit}) would be adequate within several tens percent at least for the 2D dynamical simulation, and also we confirm that the analytic expression is still conservative claim.

%%%%%%%%%%%%%%%%%%%%%%% figures

\end{document}